\documentclass[a4paper,11pt]{article}
\usepackage{pos}
\usepackage{mathrsfs}
\usepackage{amsfonts}
\usepackage{amsmath}
\usepackage{slashed}
\usepackage{array}
\usepackage{verbatim}
\usepackage{epsfig}
\usepackage{graphicx}
\usepackage{color}
\usepackage{xcolor}

\def\bal#1\eal{\begin{align}#1\end{align}}

\title{Performance of the GPU inverters with Chroma+QUDA for various fermion actions}

\author*[a,b]{Kuan Zhang}
\author[c]{Wei Sun}
\author[a,b,d,e]{Yi-Bo Yang}
\author[c]{Ren-Qiang Zhang}

\affiliation[a]{University of Chinese Academy of Sciences, School of Physical Sciences,\\
  Beijing 100049, China}

\affiliation[b]{CAS Key Laboratory of Theoretical Physics, Institute of Theoretical Physics, Chinese Academy of Sciences,\\
Beijing 100190, China}

\affiliation[c]{Institute of High Energy Physics, Chinese Academy of Sciences, Beijing 100049, China}

\affiliation[d]{School of Fundamental Physics and Mathematical Sciences, Hangzhou Institute for Advanced Study, UCAS,\\
Hangzhou 310024, China}

\affiliation[e]{International Centre for Theoretical Physics Asia-Pacific,\\
Beijing/Hangzhou, China}

\emailAdd{zhangkuan@itp.ac.cn}
\emailAdd{ybyang@itp.ac.cn}

\abstract{We present our progress on the Chroma interfaces of the twisted-mass, HISQ (highly improved staggered quark) and overlap fermion inverters using QUDA.}

\FullConference{%
 The 38th International Symposium on Lattice Field Theory, LATTICE2021
  26th-30th July, 2021
  Zoom/Gather@Massachusetts Institute of Technology
}

\begin{document}
\maketitle

\section{Introduction}

When we put the fermion on the lattice, it is unavoidable to replace the derivative by the difference of the forward and backward shifts, and introduce the fermion doubling problem. There are various kinds of the discretized fermion action to solve or avoid the doubling problem: The simplest solution is the Wilson action which adds a second order derivative term (Wilson term); and the clover action add one more term to suppress the additional chiral symmetry breaking (A$\chi$SB) introduced by the Wilson term. The twisted-mass action multiplies a complex phase on the quark mass term of the Wilson action to achieve a better suppression on A$\chi$SB comparing to the clover action case, while adding a clover term is still beneficial to suppress the residual discretization errors. On the other hand, the staggered action maps the Dirac spinors to the lattice sites to weaken the doubling problem, while introduces the taste degree of freedom which makes the data analysis to be highly non-trivial. Finally, the overlap action (and the domain wall action as its approximation) can be considered as ultimate solution to avoid A$\chi$SB, while it can be much more expensive than the other actions. 

Even though all kinds of the actions should approach the same continuum limit, their discretization error can be quite different. Thus it is essential to compare the results with different action at several lattice spacings, and have a trade-off between the statistical uncertainty and the systematic ones.  But most of the Lattice QCD software concentrates on one or two fermion actions only, and then it can be quite non-trivial to switch the actions in the production and/or compare the corresponding results. Thus it is very helpful if one can calculate fermion propagators for all the actions in a given software, like Chroma.

The Chroma~\cite{Edwards:2004sx} package is an open-source Lattice QCD software at the level 4 of the USQCD SciDAC modules, and targets an uniform interface of the various algorithms like the fermion and gauge actions, solvers, and even monomial in HMC. At the same time, the QUDA~\cite{Clark:2009wm,Babich:2011np,Clark:2016rdz} package provides the GPU-accelerated inverter for most of the fermion action except the overlap action (but most of the needed linear algebra operations are ready). Thus the missing pieces are just a Chroma interface to call kinds of the QUDA inverter, and also an implementation of the overlap action.

\begin{table*}[htbp]
  \centering
  \begin{tabular}{|c||ccrc|cc|}
\hline
Tag &  $6/g^2$ & $L$ & $T$ & $a(\mathrm{fm})$ & $m^{\textrm{w}}_{q}a$ & $c_{\mathrm{sw}}$ \\
\hline
MILC12 &  3.60 & 24 & 64 & 0.1213(9)    & -0.0695& 1.0509\\
\hline
MILC09 &  3.78 & 32 & 96 & 0.0882(7)  & -0.0514& 1.0424 \\
\hline
MILC06 &  4.03 & 48 & 144 & 0.0574(5)  & -0.0398& 1.0349 \\
\hline
\end{tabular}
  \begin{tabular}{|c||ccrc|c|}
\hline
tag &  $6/g^2$ & $L$ & $T$ & $a(\mathrm{fm})$ & $m^{\textrm{ov}}_{q}a$\\
\hline
RBC11 & 2.13 & 24 & 64 & 0.1105(3) & 0.015\\
\hline
RBC08 & 2.25 & 32 & 64 & 0.0828(3) & 0.011\\
\hline
  \end{tabular}
  \caption{Informations of the MILC and RBC ensembles used in this work.}
  \label{tab:lattice}
\end{table*}

\section{Numerical setup and results}

In this proceeding, we present the performance of GPU-accelerated inverter based on QUDA for three actions: Twisted-mass, HISQ and also overlap. The information of the gauge ensembles we used are summarized in Table~\ref{tab:lattice}. The node we used in this work include 32 CPU cores at 2 GHz and 4 V100 GPUs.

\subsection{Twisted-mass fermion}

The twisted-mass fermion action is defined as the following
\bal
S^{\rm tw}=\sum_{x,y} \bar{\psi}(x)D_W(x,y;m_{\rm cri}+\omega)\psi(x),
\eal
where 
\bal
D_w(x,y;m)=\frac{1}{2}\sum_{\mu=1,...,4,\eta=\pm}(1+\eta\gamma_{\mu})U_{\mu}(x,x+\eta\hat{n}_{\mu}a)\delta_{y,x+\eta\hat{n}_{\mu}a}-(4+m)\delta_{x,y}
\eal
is the discretized $\slashed{D}$ of the Wilson action, $m_{\rm cri}$ is the quark mass parameter to make the corresponding pion mass to be zero, and $\omega$ is a complex number and corresponds to the degenerated twisted-mass parameters. The standard Wilson action corresponds to the case with a real value of $\omega$, and a purely imaginary $\omega$ will have an automatic ${\cal O}(a)$ improvement and avoid the exceptional condition due to the instability of the critical point. One can also add a clover term on the above action to get twisted-mass clover fermion action, 
\bal
S^{\rm twc}=S^{\rm tw}+c^{\rm sw} \sigma_{\mu\nu}F^{\mu\nu}
\eal
which can further suppress the ${\cal O}(a^2)$ discretization errors.

\begin{table*}[htbp]
  \centering
  \begin{tabular}{|l||ccc|rc|}
\hline
Tag &  Ensembles & $m_{\rm cri}+\omega$ & Nodes & Invertion & Setup \\
\hline
CPU with BICGSTAB &  MILC06 & -0.0398 & 9 & 6992s & -\\
\hline
GPU with GCR &  MILC06 & -0.0398 & 9 & 652s & -\\
\hline
GPU with multigrid &  MILC06 & -0.0398 & 9 & 99s & 515s\\
\hline
GPU with multigrid &  MILC06 & -0.044+0.005i & 9 & 78s & 154s\\
\hline
  \end{tabular}
  \caption{The inverstion and setup time needed by a 12-column propagator in different cases with a similar residual $10^{-6}$.}
  \label{tab:wilson}
\end{table*}

The QUDA interface of the twisted-mass action is quite similar to that of the clover one. The only subtle issue is that the inverse of the clover term has a complex diagonal part and then can not be packed for QUDA normally. Thus one would like to enable the dynamical-clover flag in QUDA, and calculate the entire clover term (and its inverse) in QUDA directly.

The comparison of the time needed by a full propagator with 12 columns are summarized in Table~\ref{tab:wilson}. Both the pion mass in the Clover and Twisted+clover cases are tuned to be about 300 MeV. One can see that the standard GPU inverter with GCR algorithm is around 10 times further than the CPU one with BICGSTAB algorithm, and the multi-grid inverter can be even faster, with the cost of the reusable subspace setup. Comparing the clover fermion action, the time need by the twisted-mass action is shorter with similar multi-grid parameters, especially during the setup.

\subsection{HISQ fermion}

Another solution of the fermion doubling problem is the staggered fermion. With a redefinition on the fermion field, we can obtain the staggered fermion action as the following,
\bal
S^{\rm st}=\sum_{x}\bar{\psi}^{\rm st}(x)\big[\frac{1}{2}\sum_{\mu=1,...,4,\eta=\pm}\eta\gamma_{\mu}U_{\mu}(x,x+\eta\hat{n}_{\mu}a)\psi(x+\eta\hat{n}_{\mu}a)-m\big]\psi^{\rm st}(x)
\eal
where $\psi^{\rm st}(x)=\gamma_4^{x_4}\gamma_1^{x_1} \gamma_2^{x_2} \gamma_3^{x_3}\psi(x)$ at the site $x=\{x_1,x_2,x_3,x_4\}$ in the MILC conversion. Note that there is still 4 degrees of freedom which called taste, and the data analsysis with the taste mixing can be much more complicated. The HISQ action is an improved staggered action which include both the 1-step fat link (with certain smearing) and also 3-step long link~\cite{Follana:2006rc}.

One should be careful to compare the propagagor from the QUDA with that from the native Chroma HISQ inverter, since the Chroma HISQ action  uses the CPS conversion $\psi^{\rm cps}(x)=\gamma_1^{x_1} \gamma_2^{x_2} \gamma_3^{x_3}\gamma_4^{x_4}\psi(x)$, and has different sign on the mass term. Thus effectively two conversions can be related with the following relation: $\psi^{\rm Chroma}(x)=(-1)^{(t\%2)((x+y+z)\%2)+(x+y+z+t)\%2}\psi^{\rm QUDA}$. 

Similarly, we can see significant speed up of the GPU inverter comparing to the CPU one, especially when we use fewer nodes to do the test on small lattices. For multigrid, we apply KD Preconditioning~\cite{Brower:2018ymy}. 

The multigrid inverter requires very long time to generate the subspace, and the inverter is not faster than the standard CG algrithm even after the subspace is generated. The parameters we use are shown in Fig.1.

\begin{figure}[tbph]
  \centering
  \includegraphics[width=16cm]{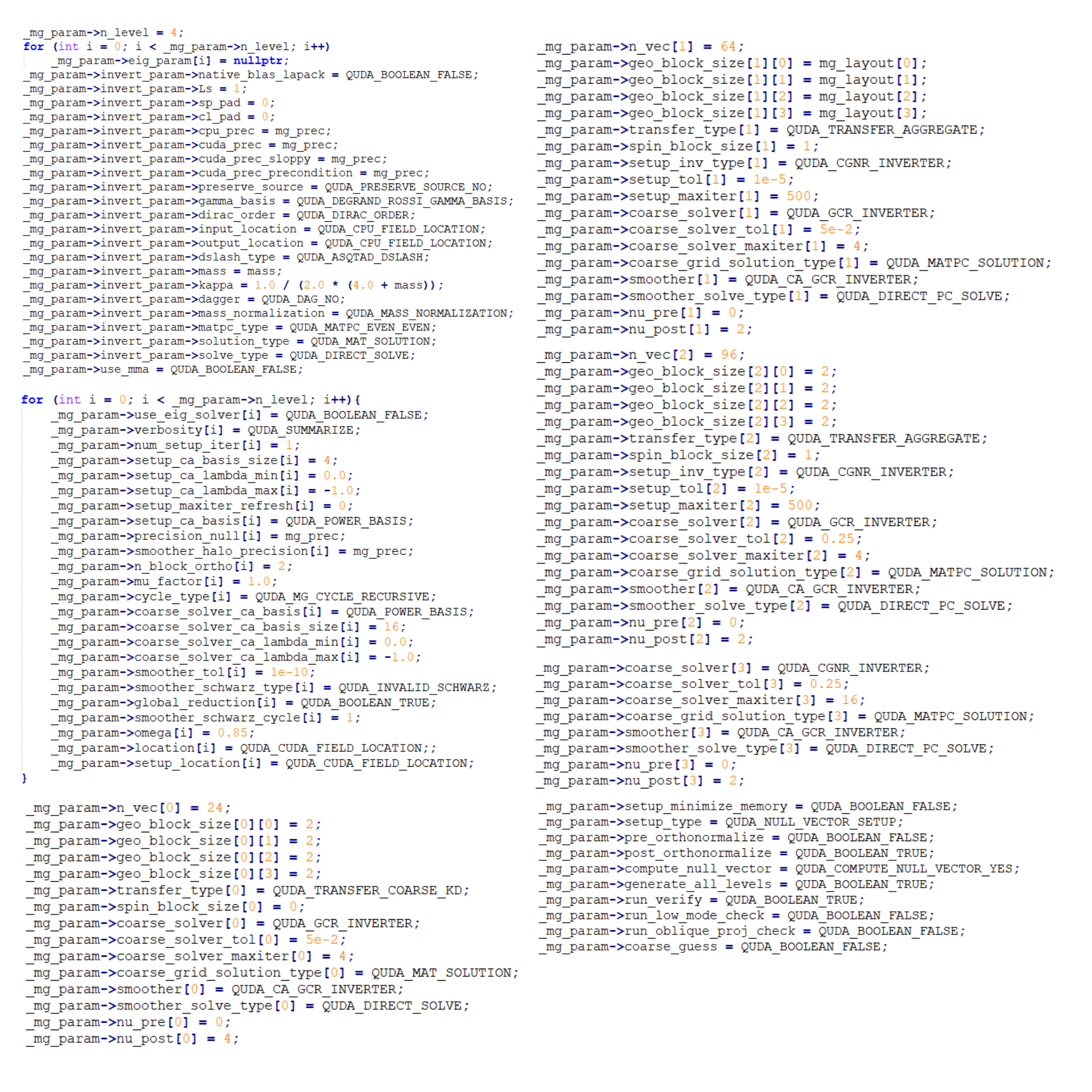}
  \caption{The parameters for HISQ multigrid, most of which are copied from MILC interface.}
  \label{fig:HISQ multigrid}
  \end{figure}

\begin{table*}[htbp]
  \centering
  \begin{tabular}{|l||ccc|cc|}
\hline
tag &  Ensembles & $m$ & Nodes & Inverter & Setup \\
\hline
CPU with CG &  MILC12 & 0.0102 & 1 & 523s & -\\
\hline
GPU with CG &  MILC12 & 0.0102 & 1 & 17s & -\\
\hline
GPU with multigrid &  MILC12 & 0.0102 & 1 & 26s & 334s\\
\hline
CPU with CG &  MILC09 & 0.0074 & 3 & 1086s & -\\
\hline
GPU with CG &  MILC09 & 0.0074 & 3 & 23s & -\\
\hline
GPU with multigrid &  MILC09 & 0.0074 & 3 & 42s & 311s\\
\hline
  \end{tabular}
  \caption{The time needed by a 3-column HISQ propagator with either CPU or GPU inverter, on two ensembles.}
  \label{tab:hisq}
\end{table*}

\subsection{Overlap fermion}

\begin{figure}[tbph]
  \centering
  \includegraphics[width=10cm]{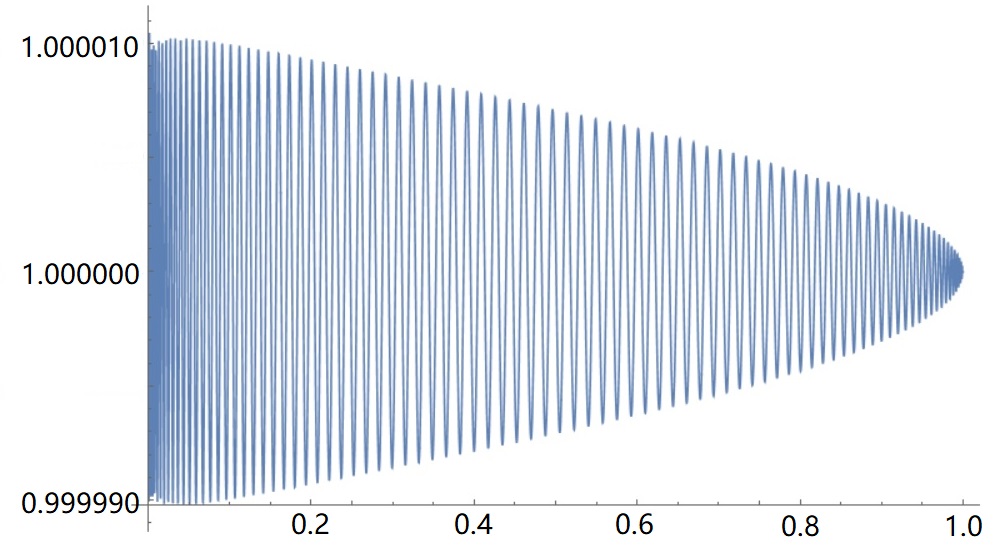}
  \caption{The sign function approximated by a Chebyshev polynomial with no more than $10^{-5}$ deviation.}
  \label{fig:chebyshev polynomial}
  \end{figure}

The solution to avoid the entire fermion doubling problem is the chiral fermion satisfying the Ginsparg-Wilson relation, likes the overlap fermion,
\bal
S^{\rm ov}=\sum_x \bar{\psi}(x)D_{\rm ov}(x,y)\psi(y),
\eal
where $D_{ov}=\frac{1}{\rho}(1+\frac{D_w(-\rho)}{\sqrt{D^{\dagger}_w(-\rho)D_w(-\rho)}})$ with $\rho\sim 1.5$. The term $\frac{D_w(-\rho)}{\sqrt{D^{\dagger}_w(-\rho)D_w(-\rho)}}$ can be rewritten into $\gamma_5\epsilon(\gamma_5D_w(-\rho))$, where $\epsilon(x)$ is the sign function. Usually, we solve the smallest $\cal O$(100-1000) eigenvectors of $\gamma_5D_w(-\rho)$ at the accuracy $10^{-12}$ and obtain the sign function of this subspace explicitly, and use the Chebyshev polynomial likes what shown in Fig.~\ref{fig:chebyshev polynomial} to approximate the sign function of $\gamma_5D_w(-\rho)$ in the other subspace with the larger eigenvalues. 

For the Chebyshev polynomial, we estimate the ranks with an empirical formula, set several initial $x$ values and solve the equations of the coefficients to make the sign function at those values to be exact. Of course the residual will not vanish at the other points, and we need to repeat the procedure at the extreme points of the new polynomials until the precision goal is reached at the new extreme points. Note that one should use use the Clenshaw recursion to define the Chebyshev polynomial to suppress the round-off error.

In order to enhance the contribution from the low mode, the Chebyshev acceleration is also used in the eigenvector solver. Note that the polynomial will change the eigenvalues of a matrix, but its eigenvectors are kept unchanged.

One can apply the polynomial of a dslash several times on a random vector $b$ and obtain its Krylov array.
\bal
{b,Ab,...,A^{k-1}b},
\eal
For Arnoldi algorithm, we can get a Heisenberg matrix after the Schmidt orthogonalization
\bal
AQ = \begin{pmatrix} Aq_1 & Aq_2 & \cdots & Aq_k\end{pmatrix} = \begin{pmatrix} q_1 & q_2 & \cdots & q_k\end{pmatrix} \begin{pmatrix} h_{1,1} & h_{1,2} & h_{1,3} & \cdots & h_{1,k}\\ h_{2,1} & h_{2,2} & h_{2,3} & \cdots & h_{2,k} \\ 0 & h_{3,2} & h_{3,3} & \cdots & h_{3,k} \\ \vdots & \ddots & \ddots & \ddots & \vdots \\ 0 & \cdots & 0 & h_{k,k-1} & h_{k,k} \end{pmatrix}.
\eal
Eventually the eigenvectors can be obtained by diagonalizing the Heisenberg matrix. Note that we can use the special QR factorization for Heisenberg matrix to suppress the round-off error. When the amount of the eigenvector is large, the restart algorithm is essential to save the GPU memory by the decrease of extra space. 

 As shown in Table~\ref{tab:ov1}, the GPU eigensolver can be much faster than the CPU one, on both the ensembles. Our codes use the restarted Arnoldi algorithm which is similar to the algorithm in the GWU-code.  

\begin{table*}[htbp]
  \centering
  \begin{tabular}{|c||ccc|cc|}
\hline
tag &  Ensembles & Number of eigenvectors & Nodes & time(Chroma) & time(GWU-code)\\
\hline
CPU &  RBC11 & 200 & 1 & 8070s & -\\
\hline
GPU &  RBC11 & 200 & 1 & 180s & 225s\\
\hline
CPU &  RBC08 & 200 & 4 & 4704s & -\\
\hline
GPU &  RBC08 & 200 & 4 & 281s & 317s\\
\hline
  \end{tabular}
  \caption{The time needed to generate the 200 eigenvalues of $\gamma_5D_{w}$ with residual $10^{-15}$.}
  \label{tab:ov1}
\end{table*}

\begin{table*}[htbp]
  \centering
  \begin{tabular}{|c||ccc|cc|}
\hline
tag &  Ensembles & Number of eigenvectors & Nodes & time(Chroma) & time(GWU-code) \\
\hline
CPU &  RBC11 & 200 & 1 & >12h & -\\
\hline
GPU &  RBC11 & 200 & 1 & 7384s & 16110s\\
\hline
CPU &  RBC08 & 200 & 4 & >12h & -\\
\hline
GPU &  RBC08 & 200 & 4 & 8893s & 14370s\\
\hline
  \end{tabular}
  \caption{The time needed to generate the 200 eigenvalues of $D^{L/R}_{ov}$ with residual $10^{-12}$.}
  \label{tab:ov2}
\end{table*}

Similarly we can solve the low lying eigenvectors of the $D_{ov}$ to accelerate the inversion of the overlap propagator. Since the $D_{ov}$ is so-called gamma5 Hermite, we considered the projected $D^{L/R}_{ov}=(1\pm \gamma_5)D_{ov}(1\pm\gamma_5)$ to make sure that the eigenvalues are real for the convergence of the Arnoldi eigenslover and just solve the eigenvectors in the chiral sector with zero modes. In the end, we reconstruct the full spinor.

As in Table~\ref{tab:ov2}, the overlap eigensolver based on QUDA can be faster than our previous one using the GWU-code~\cite{Alexandru:2011ee,Alexandru:2011sc}. On the RBC11 ensemble, 50 $D_w$ operations take 0.088s in GWU-code while just 0.025s in QUDA with 1 node, as we combine the $\gamma_5$ and $D_w$ operations into one kernel to save the bandwidth in the QUDA code, and QUDA can take advantage from its auto-tuning. But the cost of the vector operations are similar in both QUDA and GWU-code, thus the difference in the eigensolver performance is smaller, especially in the $H_w$ one which spends fewer time in the matrix operations. As a larger scale, the RBC08 ensemble uses 4 nodes and then the network has much stronger impact on the performance, so the $D_w$ performance in two cases are relatively closer, as 50 $D_w$ take 0.148s in GWU-code and 0.087s in QUDA.

We also implemented the deflation~\cite{xQCD:2010pnl} and multi-mass~\cite{Jegerlehner:1996pm} algorithm for the overlap propagator in Chroma to take the advantage of the overlap fermion definition. On the RBC11 ensembles, we generate overlap propagators with the mass of 0.03, 0.05 and 0.10 within the residual 1e-8. It takes 1150s in GWU-code while just 698s in QUDA with 1 node for the calculations. The RBC08 ensembles use 4 nodes and the speed up of the propagator solver is similar to that of $D_{ov}$ eigensolver. We choose the mass of the propagators as 0.03, 0.05 and 0.10 and the tolerance is set to be 1e-8. The inversion takes 1133s in GWU-code and 903s in QUDA. The performance in two cases are also closer.

\section{Summary}

In summary, we write the Chroma interfaces of the QUDA twist-mass and HISQ inverters, and implemented the overlap fermion eigensolver and inverter based on the QUDA dslash kernel and linear algebra operations. It turns out that the QUDA can provide significant speed up on the above three actions with the uniform Chroma interface, while that the HISQ multigrid solver would not be properly tuned and then require further efforts. It paves the way to compare the statistical and systematic uncertainties of the same physical observable with different actions, with an uniform environment. 

\section*{Acknowledgement}

We thank the MILC and RBC/UKQCD collaborations for providing us their gauge configurations, and Ke-Long Zhang and Long-Cheng Gui, for useful information and discussion. The calculations were performed using the Chroma software suite~\cite{Edwards:2004sx} with QUDA~\cite{Clark:2009wm,Babich:2011np,Clark:2016rdz} and GWU-code~\cite{Alexandru:2011ee,Alexandru:2011sc} through HIP programming model~\cite{Bi:2020wpt}. 

\bibliographystyle{unsrt}
\bibliography{ref}

\begin{thebibliography}{10}

\bibitem{xQCD:2010pnl}
A.~Li et~al.
\newblock {Overlap Valence on 2+1 Flavor Domain Wall Fermion Configurations
  with Deflation and Low-mode Substitution}.
\newblock {\em Phys. Rev. D}, 82:114501, 2010.

\bibitem{Jegerlehner:1996pm}
Beat Jegerlehner.
\newblock {Krylov space solvers for shifted linear systems}.
\newblock 12 1996.

\bibitem{Follana:2006rc}
E.~Follana, Q.~Mason, C.~Davies, K.~Hornbostel, G.~P. Lepage, J.~Shigemitsu,
  H.~Trottier, and K.~Wong.
\newblock {Highly improved staggered quarks on the lattice, with applications
  to charm physics}.
\newblock {\em Phys. Rev. D}, 75:054502, 2007.

\bibitem{Edwards:2004sx}
Robert~G. Edwards and Balint Joo.
\newblock {The Chroma software system for lattice QCD}.
\newblock {\em Nucl. Phys. Proc. Suppl.}, 140:832, 2005.
\newblock [,832(2004)].

\bibitem{Clark:2016rdz}
M.~A. Clark, Blint Jo, Alexei Strelchenko, Michael Cheng, Arjun Gambhir, and
  Richard Brower.
\newblock {Accelerating Lattice QCD Multigrid on GPUs Using Fine-Grained
  Parallelization}.
\newblock 2016.

\bibitem{Bi:2020wpt}
Yu-Jiang Bi, Yi~Xiao, Ming Gong, Wei-Yi Guo, Peng Sun, Shun Xu, and Yi-Bo Yang.
\newblock {Lattice QCD package GWU-code and QUDA with HIP}.
\newblock {\em PoS}, LATTICE2019:286, 2020.

\bibitem{Clark:2009wm}
M.~A. Clark, R.~Babich, K.~Barros, R.~C. Brower, and C.~Rebbi.
\newblock {Solving Lattice QCD systems of equations using mixed precision
  solvers on GPUs}.
\newblock {\em Comput. Phys. Commun.}, 181:1517--1528, 2010.

\bibitem{Babich:2011np}
R.~Babich, M.~A. Clark, B.~Joo, G.~Shi, R.~C. Brower, and S.~Gottlieb.
\newblock {Scaling Lattice QCD beyond 100 GPUs}.
\newblock In {\em {SC11 International Conference for High Performance
  Computing, Networking, Storage and Analysis Seattle, Washington, November
  12-18, 2011}}, 2011.

\bibitem{Alexandru:2011ee}
A.~Alexandru, C.~Pelissier, B.~Gamari, and F.~Lee.
\newblock {Multi-mass solvers for lattice QCD on GPUs}.
\newblock {\em J. Comput. Phys.}, 231:1866--1878, 2012.

\bibitem{Alexandru:2011sc}
Andrei Alexandru, Michael Lujan, Craig Pelissier, Ben Gamari, and Frank~X. Lee.
\newblock {Efficient implementation of the overlap operator on multi-GPUs}.
\newblock In {\em {Proceedings, 2011 Symposium on Application Accelerators in
  High-Performance Computing (SAAHPC'11): Knoxville, Tennessee, July 19-20,
  2011}}, pages 123--130, 2011.

\bibitem{Brower:2018ymy}
Richard~C. Brower, M.~A. Clark, Alexei Strelchenko, and Evan Weinberg.
\newblock {Multigrid algorithm for staggered lattice fermions}.
\newblock {\em Phys. Rev. D}, 97(11):114513, 2018.

\end{thebibliography}

\end{document}